\documentclass{elsart}

\usepackage[english]{babel}
\usepackage[latin1]{inputenc}
\usepackage{amssymb}
\usepackage{mathrsfs}
\usepackage{amsmath}
\usepackage[dvips]{graphicx}
\usepackage{epsfig,bm}

\begin{document}
\begin{frontmatter}
\title{Enhancement of particle trapping in the wave-particle interaction}

\author[cpt]{R. Bachelard\thanksref{Corresponding author}},
\author[unifi]{A. Antoniazzi},
\author[cpt]{C. Chandre},
\author[unifi]{D. Fanelli}, 
\author[cpt]{M. Vittot}
\address[cpt]{Centre de Physique Th\'eorique\thanksref{cptumr}, CNRS Luminy, Case 907, F-13288 Marseille Cedex 9, France}
\address[unifi]{Dipartimento di Energetica and CSDC, Università di Firenze, INFN, via S. Marta, 3, 50139 Firenze, Italy}
\thanks[Corresponding author]{Tel.~: +33 4 91 26 95 47 ; Fax~: +33 4 91 26 95 53\\
Email~: bachelard@cpt.univ-mrs.fr}
\thanks[cptumr]{Unit\'e Mixte de Recherche (UMR 6207) du CNRS, et des universit\'es Aix-Marseille I, Aix-Marseille II et du Sud Toulon-Var. Laboratoire affili\'e \`a la FRUMAM (FR 2291).}

\begin{abstract}
The saturated dynamics of a Single Pass Free Electron Laser is considered within a simplified mean field approach.
A method is proposed to increase the size of the {\it macro-particle}, which is responsible for the oscillations of the intensity of the wave. This approach is based on the reconstruction of invariant tori of the dynamics of test particles. To this aim a dedicated control term is derived, the latter acting as a	small apt perturbation of the system dynamics. Implications of these findings are discussed in relation to the optimization of the laser source.
{\it PACS~:} 41.60.Cr, 47.10.Df, 05.45.Gg
\end{abstract}

\begin{keyword}
~Free Electron Laser \sep Hamiltonian dynamics \sep chaos control 
\end{keyword}
\end{frontmatter}

\section*{Introduction}
A Free Electron Laser (FEL) generates a tunable, coherent, high power
radiation. A FEL differs from conventional lasers in
using a relativistic electron beam as its lasing medium. The physical
mechanism responsible of the light emission and amplification is the
interaction between a relativistic electron beam and a magnetostatic
periodic field generated in the undulator. Due to the effect of the magnetic  
field, the electrons are forced to follow sinusoidal trajectories, thus emitting 
synchrotron radiation. This initial seed, termed {\it spontaneous emission}, is 
then amplified along the undulator until the laser effect is reached.

\par

Among different schemes, Single-Pass high-gain FELs are particularly attractive since they hold the 
promise of resulting in flexible light sources of broad theoretical and applied interests.  
The coupled evolution of the radiation field and the $N$ particle beam
in a Single-Pass FEL can be successfully modeled within the framework of a simplified Hamiltonian picture
\cite{bonifacio}. The $N+1$ degree of freedom Hamiltonian 
displays a kinetic contribution, associated with the particles, and a potential term accounting for the 
self-consistent coupling between the particles and the field. Hence, direct inter-particles interactions 
are neglected, even though an effective coupling is indirectly provided because of the interaction with the wave.

The theory predicts a linear exponential instability and a late oscillating
saturation, for the amplitude of the radiation field. Inspection of the asymptotic phase-space
suggests that a bunch of particles gets trapped in the resonance and forms a clump
that evolves as a single macro-particle localized in space.
The  remaining particles are almost uniformly distributed between two
oscillating boundaries, and populate the so called {\it chaotic sea}.

This observation allowed to derive a simplified Hamiltonian model to
characterize the saturated evolution of the laser. Such reduced
formulation consists in only four degrees of freedom, namely the wave, the macro-particle and the
two boundaries, delimiting the portion of space occupied by the
the uniform halo surrounding the inner bulk \cite{tennyson,antoniazzi}. 

Furthermore, the macro-particle rotates around a well defined fixed point
and this microscopic dynamics is shown to be responsible for the macroscopic 
oscillations observed at the intensity level. It can be therefore hypothesized that a significant 
reduction in the intensity fluctuations can be gained by implementing a 
dedicated control strategy, aimed at confining the macro-particle in space.
In addition, the size of the macro-particle is directly related to the bunching
parameter, a quantity of paramount importance in FEL context \cite{antoniazzi}.

Smith {\it et al} \cite{smith} showed that a test-wave, being characterized by a frequency close to
the one of the wave, can effectively destroy the macro-particle and activate a consequent detrapping process.
Dimonte {\it et al} \cite{dimonte} implemented this approach on a Travelling Wave
Tube and  detected a significant reduction of the intensity oscillations, 
followed, however, by an undesired systematic collapse of its mean-value.
 
\par
In this paper, we focus on the macro-particle, spontaneously established as a result of the wave-particle interaction process in the saturated regime.
In particular we develop a dedicated technique to influence and, possibly, control its evolution, thus opening up the perspective
of defining innovative approaches aiming at stabilizing the laser signal. 
To this end, we consider a test-particle Hamiltonian in a mean-field approach
and calculate a small but appropriate {\it control term} which acts as a perturbation. The latter is shown to
induce an increase in size of the macro-particle.  A regularization of the
dynamics is also observed, as confirmed by the reconstruction of invariant
tori around the massive core.

\par
The paper is organized as follows. In Sec.~\ref{sec1}, we introduce the mean-field model and shortly outline its derivation from the
original $N$-body Hamiltonian. In Sec.~\ref{sec2}, a test-particle is ``controlled'',
through the reconstruction of invariant tori of its dynamics. Finally, we draw our conclusion and discuss possible implication of the present analysis.

\section{The mean-field model}\label{sec1}

As previously anticipated, the dynamics of a Single-Pass FEL is successfully captured by the following Hamiltonian \cite{bonifacio}~:
 \begin{equation}\label{HN}
 H_{N} = \sum_{i = 1}^{N} [ \frac{p_i^{2}}{2} - 2 \sqrt{\frac{I}{N}} \cos{(\theta_i + \phi)} ],
\end{equation}
where $(\theta_i,p_i)$ are the position and momentum of the $i$-th particle and $(\phi,I)$ stand respectively for the phase and intensity of the radiation. 
\begin{figure}[t] \label{intensity}
\setlength\unitlength{1in}
	\begin{picture}(6,2)
    \put(0,0){\includegraphics[width=2.7in]{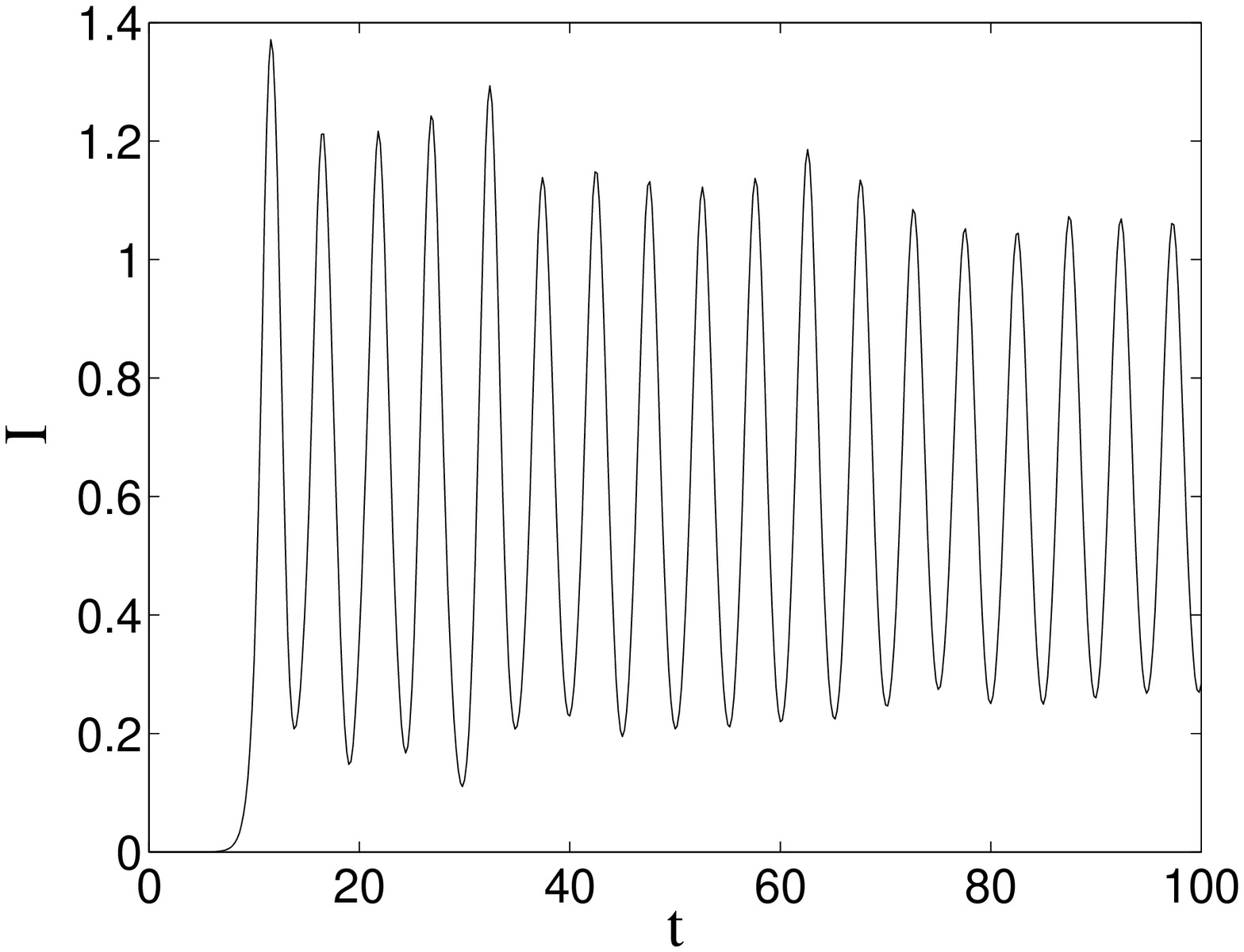}}
    \put(3.1,1.1){\includegraphics[width=1.2in]{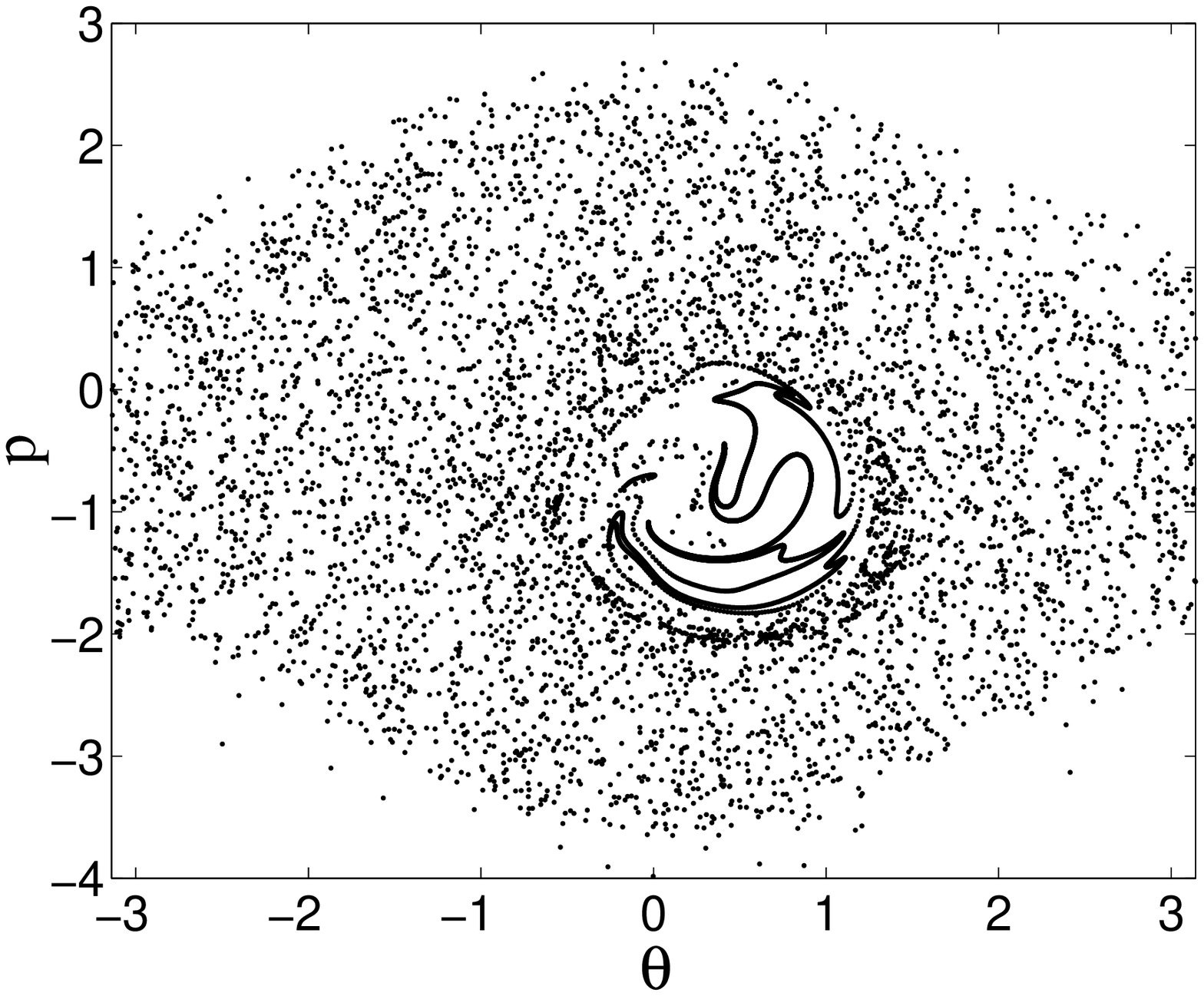}}
		\put(4.5,1.1){\includegraphics[width=1.2in]{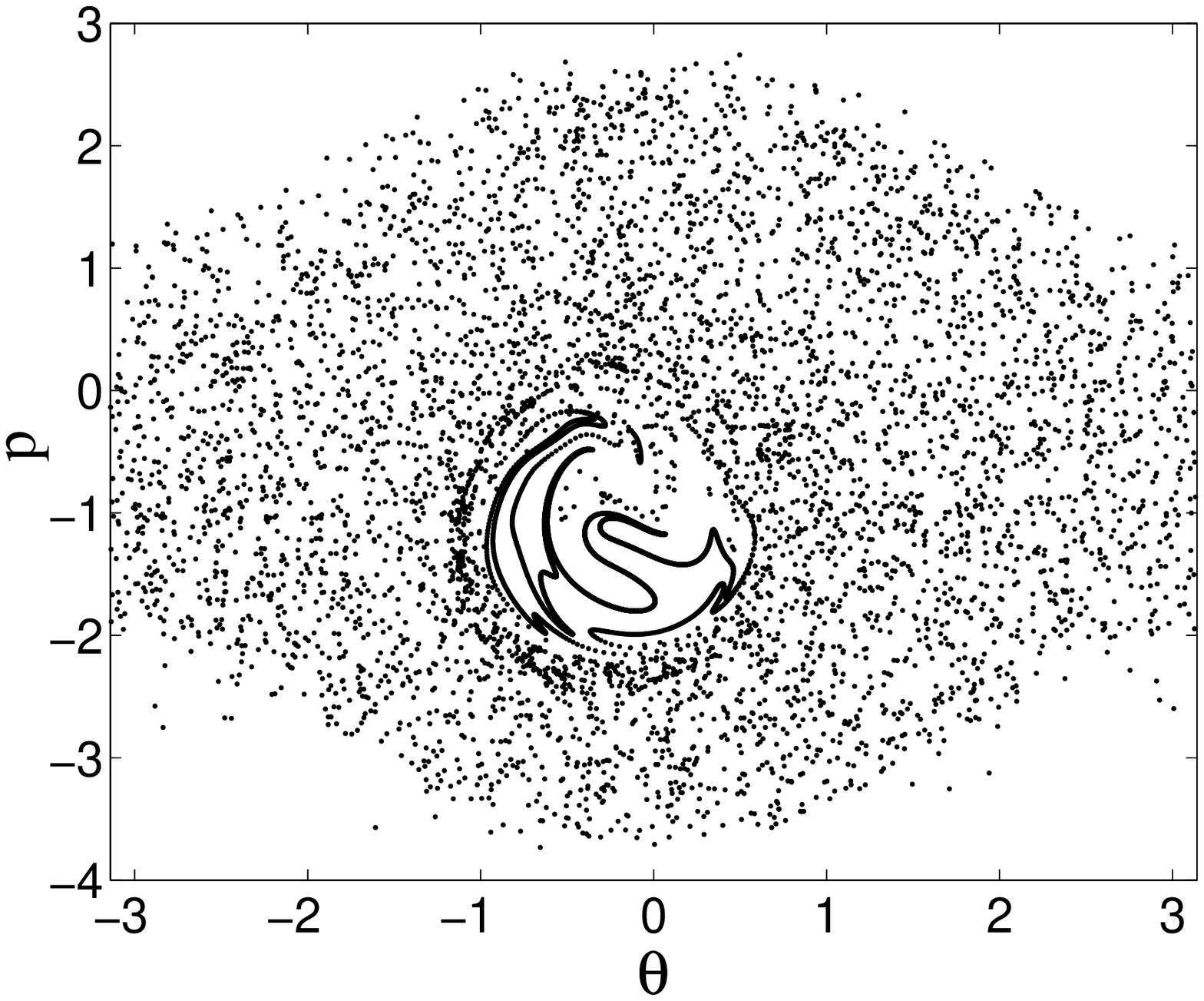}}
		\put(3.1,0.1){\includegraphics[width=1.2in]{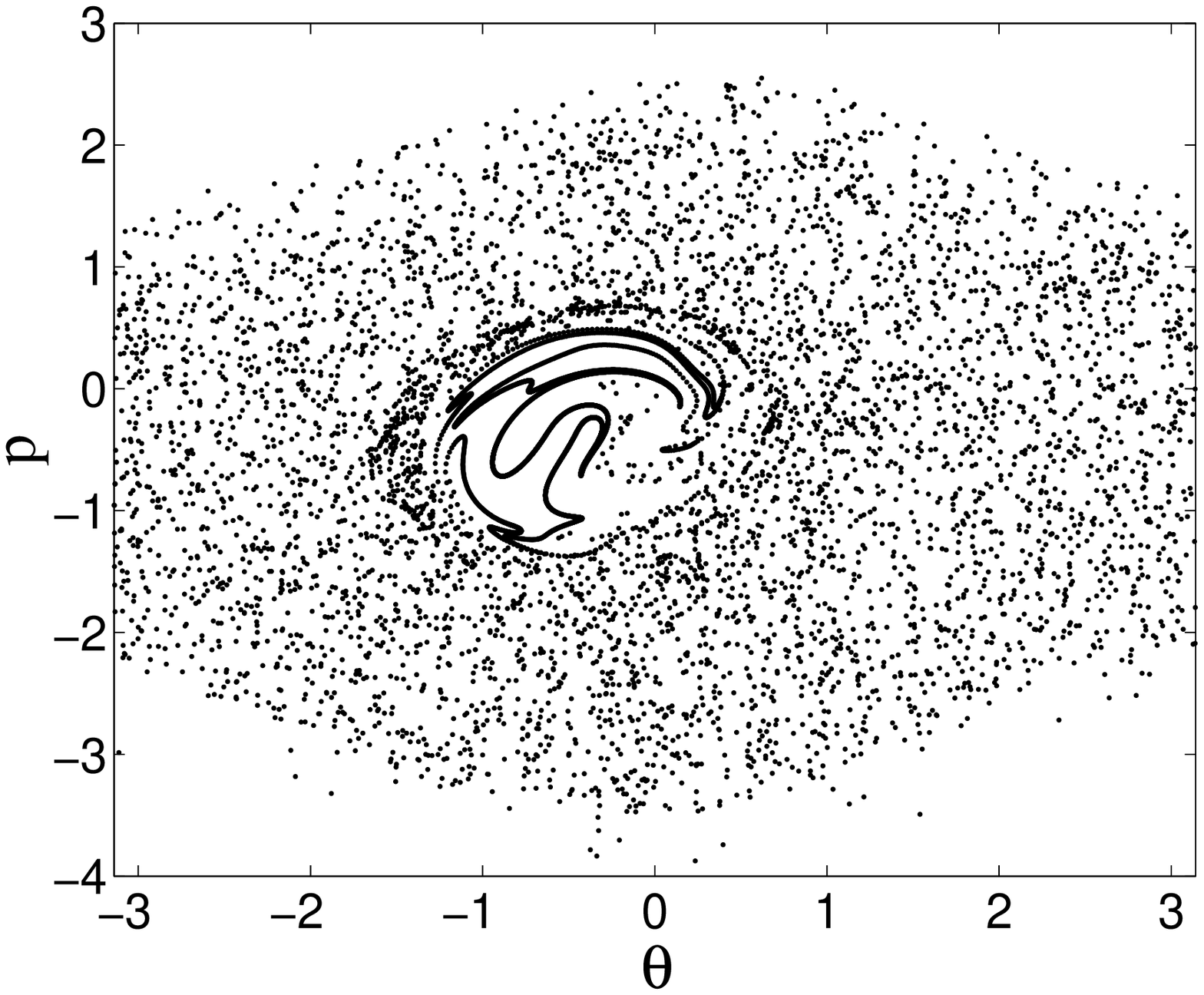}}
		\put(4.5,0.1){\includegraphics[width=1.2in]{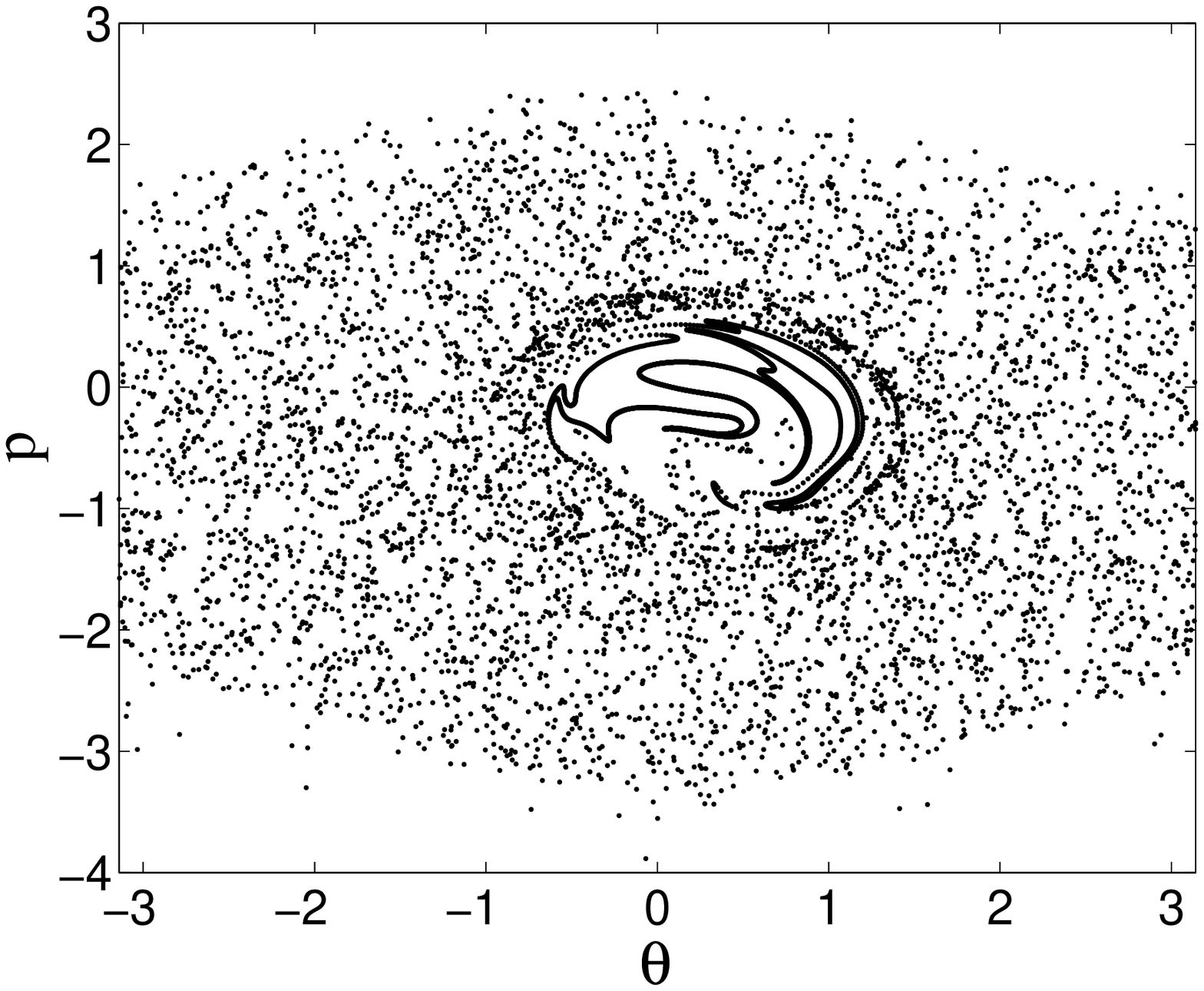}}
	\end{picture}
 \caption{Left~: Normalized intensity of the FEL's radiation simulated from Hamiltonian (\ref{HN}). Right~: $N=10000$ electrons in phase space, at $t=400$, $401.25$, $402.5$ and $403.75$.}
 \end{figure}
In the mean-field model, the conjugated variables $(\phi,I)$ are replaced by two functions of time $\phi(t)$ and $I(t)$, obtained from 
direct simultations of the self-consistent dynamics. In other words, this amounts to formally neglecting the action of the electrons on the field. 

The $N$-body Hamiltonian (\ref{HN}) can therefore be mapped into
\begin{equation*}
\tilde{H_{N}} = \sum_{i = 1}^{N} H_{1p} (\theta_i,p_i,t),
\end{equation*}
where 
 \begin{equation}\label{H15}
H_{1p} (\theta,p,t) = \frac{p^{2}}{2} - 2 \sqrt{\frac{I(t)}{N}} \cos{(\theta + \phi(t))}.
\end{equation} 
Hence, the dynamics of a FEL can be adressed by studying the evolution  of a test particle, obeying Hamiltonian ($\ref{H15}$). 
The functions $I(t)$ and $\phi(t)$ act as external fields and are here imposed by assuming their simplified
asymptotic behaviour as obtained by a frequency analysis \cite{laskar}~:
\begin{equation*} 2 \sqrt{\frac{I(t)}{N}} e^{i \phi(t)} \approx F - \epsilon \sum_{k=1}^{K} W_k e^{i \omega_k t},
\end{equation*}
in the reference frame of the wave.

\section{Hamiltonian control of a test-particle}\label{sec2}
 The Hamiltonian control addresses systems which are close to integrable, i.e. whose Hamiltonian can be written as $H=H_0+V$, where $H_0$ is integrable 
 and $V$ a perturbation of order $\epsilon$ (compared to $H_0$). The results we use here have been proven rigorously \cite{vittot04,vittot05}. 
 In practice, it can be shown that a suitable {\it control term} $f$ of order $\epsilon^2$ exists such that $H_0 +V+f$ has an invariant torus at a given frequency 
 $\omega_0$. In our case, the perturbation corresponds to the oscillating part of the intensity. The interaction term of Hamiltonian (\ref{H15}) reads
\begin{equation*} 2\sqrt{\frac{I(t)}{N}}\cos{(\theta-\phi(t))}=F \cos{\theta}-\epsilon \rm{Re}(e^{i \theta}W(t)). \end{equation*} 
 Thus, our integrable Hamiltonian can be cast in the form
 \begin{equation}\label{H0} H_0=\frac{p^2}{2}-F \cos{\theta}, \end{equation} 
whereas the time-dependent perturbation $V$ is identified as
 \begin{equation}\label{V} V(t,\theta)=\epsilon \rm{Re}(e^{i \theta}W(t)). \end{equation}
 First, we express Hamiltonian (\ref{H0}) into action-angle variables $(\varphi,J)$ \cite{lichtenberg}. Then, we expand $H_0$ around $J=J_0$, which in turn identifies 
the region where the invariant torus is reconstructed in~:
\begin{equation*} H_0(J)=E_0+\omega_0(J-J_0)+\delta (J-J_0)^2+O((J-J_0)^3). \end{equation*}
Likewise, the $\theta$-component of perturbation (\ref{V}) is expanded as
\begin{equation*} e^{i \theta}= \sum_{m=0}^{M} \sum_{n=-L}^{L} \alpha_{m,n} (J-J_0)^m e^{i n \varphi}+O((J-J_0)^{M+1}) \end{equation*}
The control term reads \cite{vittot05}
\begin{equation*} f(\varphi,t)=V(J_0,\varphi,t)-V(J_0-\Gamma \partial_\varphi V(J_0,\varphi,t),\varphi,t), \end{equation*}
where $\Gamma$ is a linear operator acting on an element of the Fourier basis as~:
\begin{equation*} \Gamma e^{i(\omega t+n \varphi)}=\frac{e^{i(\omega t+n \varphi)}}{i(\omega + n \omega_0)}. \end{equation*}

At the second order in $\epsilon$ (which is the main term), the control term reads
\begin{equation*} f(\varphi,t)=\epsilon^2 w(\varphi,t) \Gamma \partial_\varphi v(\varphi,t) - \epsilon^2 \delta (\Gamma \partial_\varphi v(\varphi,t))^2 , \end{equation*}
where $v$ and $w$ are the first terms in the expansion of $V$~:
\begin{equation*} V(J,\varphi,t) = v(\varphi,t) + (J-J_0) w(\varphi,t) + +O((J-J_0)^{2}), \end{equation*}
which gives
\begin{equation*} \Gamma \partial_\varphi v(\varphi,t) = \sum_{k=0}^{K} \sum_{n=-L}^{L} \frac{n \alpha_{0,n} W_k}{\omega_0 n+\omega_k} e^{i(n\varphi+\omega_k t)}, \end{equation*}
and 
\begin{equation*} w(\varphi,t) = \sum_{k=0}^{K} \sum_{n=-L}^{L} \alpha_{1,n} W_k e^{i(n\varphi+\omega_k t)}, \end{equation*}
where $W_k$ are the Fourier coefficients of $W$~:
\begin{equation*} W(t)=\sum_{k=0}^{K} W_k e^{i \omega_k t}. \end{equation*}
 Numerically, two Fourier modes are considered for $W(t)$ ($K=2$), and eleven for the $\theta$-dependent part of the perturbation (\ref{V}) ($L=5$). The perturbation is of amplitude $\epsilon=1/5$. The expansion is performed around $J_0 \approx 1.33$ (which corresponds to the energy $E_0=0$).
 
 \begin{figure}[t] 
  \centerline{
    \mbox{\includegraphics[width=2.5in]{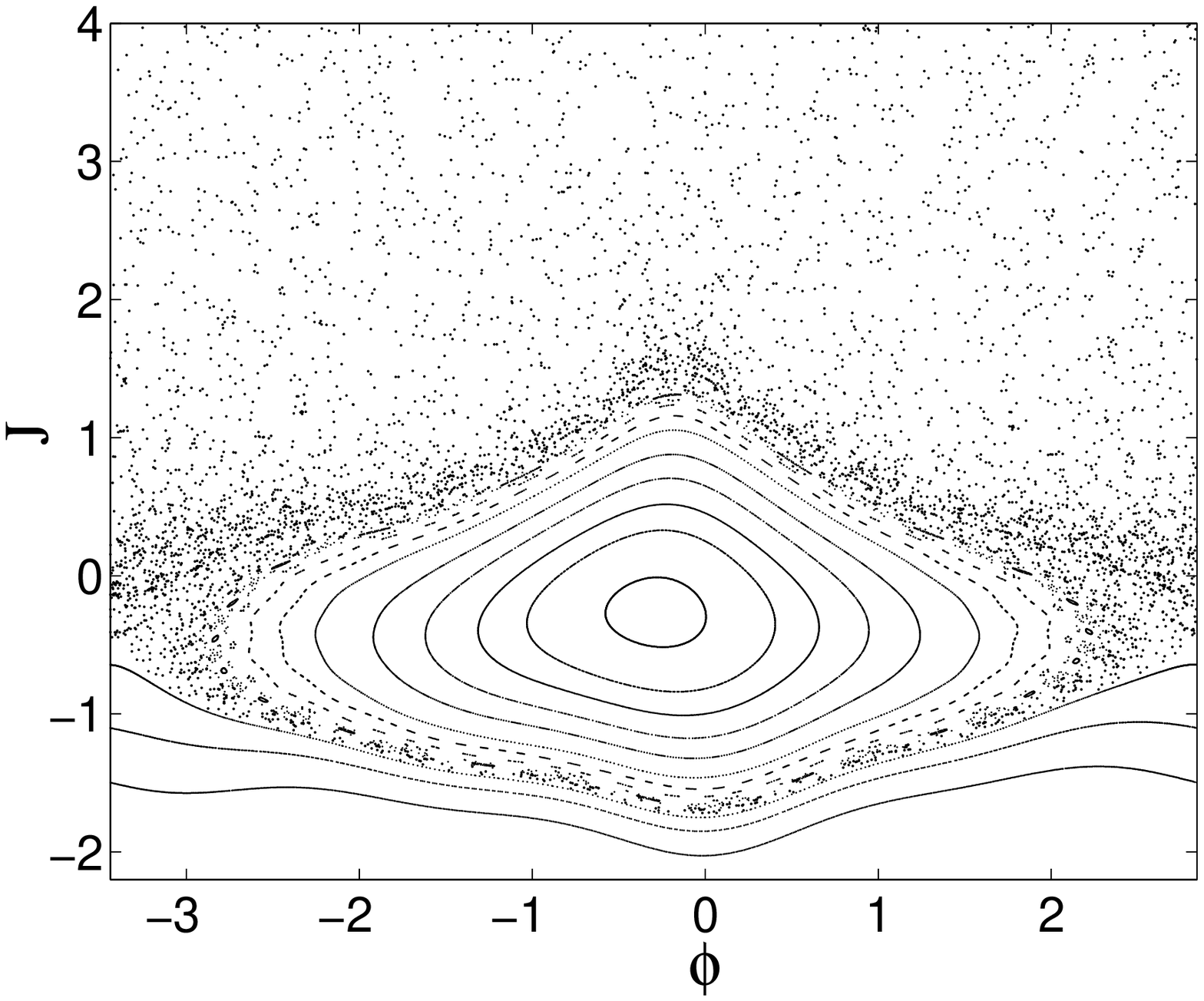}}
    \mbox{\includegraphics[width=2.5in]{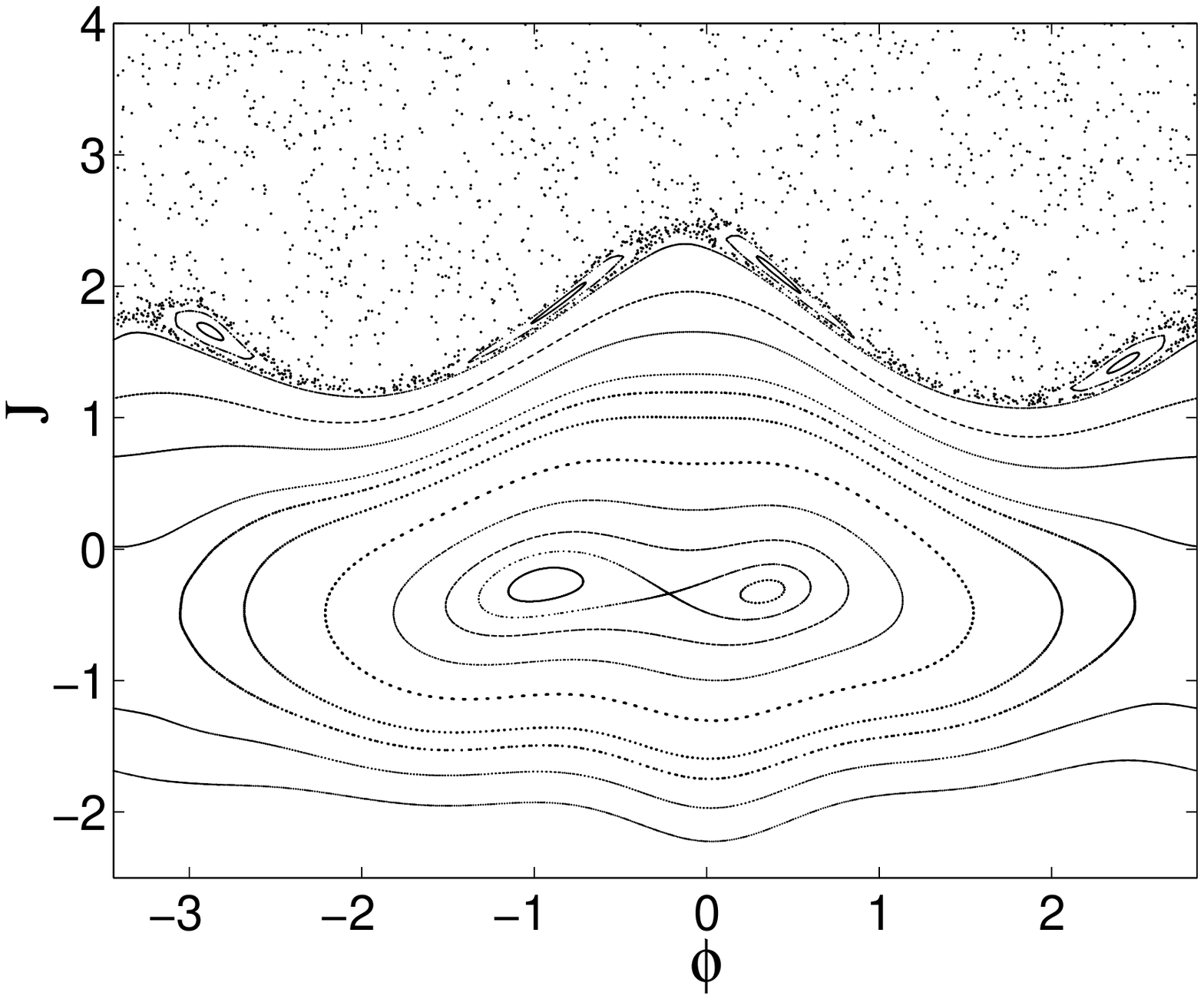}}}
  \caption{Poincaré sections of a test-particle of Hamiltonian $H_0(J)+V(J,\varphi,t)$ (left) and $H_0(J)+V(J,\varphi,t)+f(\varphi,t)$ (right), in the pendulum action-angle variables.\label{phij}}
\end{figure}

\par In action-angle variables, the regularization of the dynamics (see Fig.\ref{phij}) is clearly seen. Rather than a single torus, the control term generates a continuous set of invariant tori, thus expanding the regular domain in phase space.
\par
\begin{figure}[t] 
 \centerline{
  \mbox{\includegraphics[width=2.8in,height=2.1in]{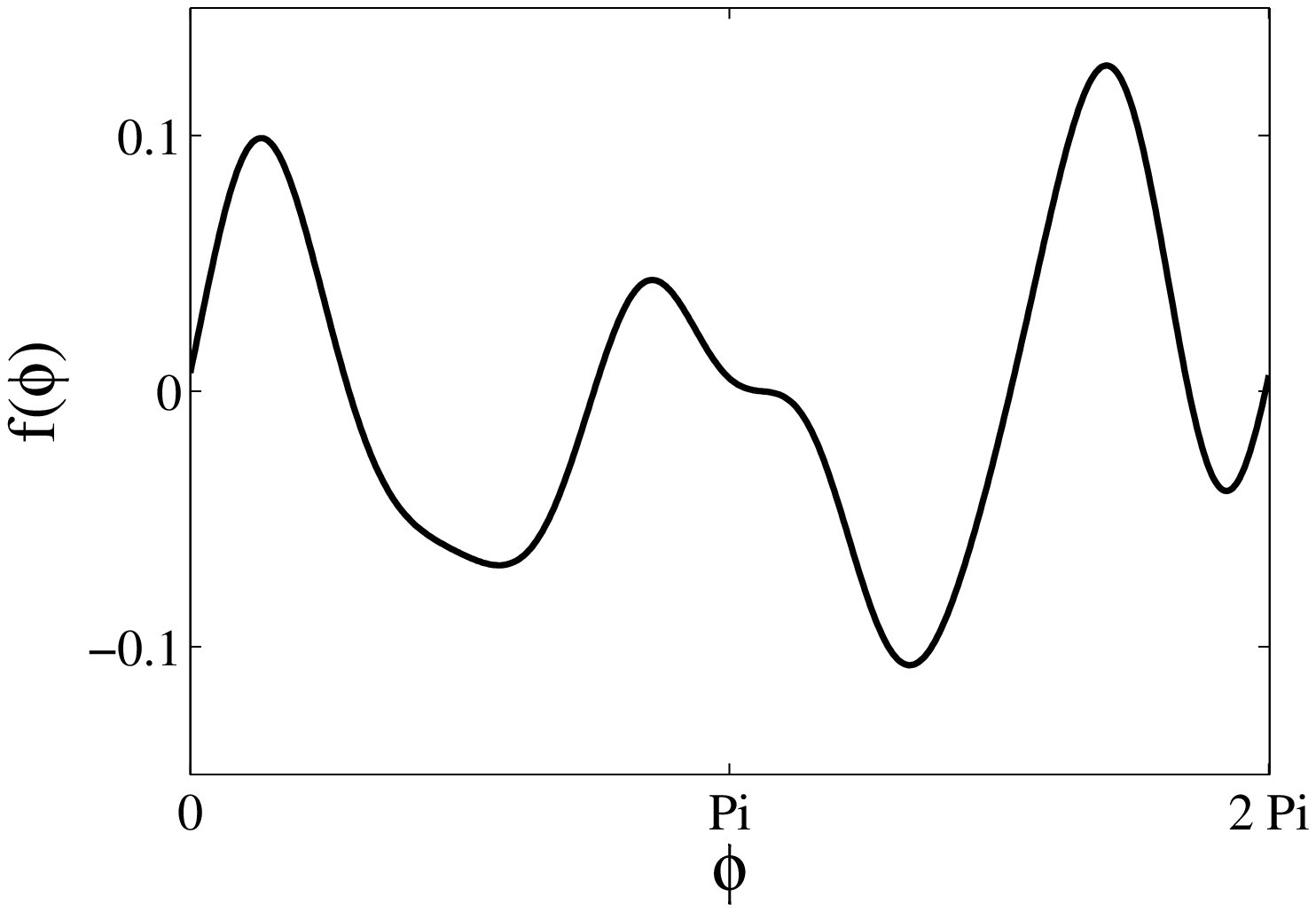}}
  \mbox{\includegraphics[width=2.5in]{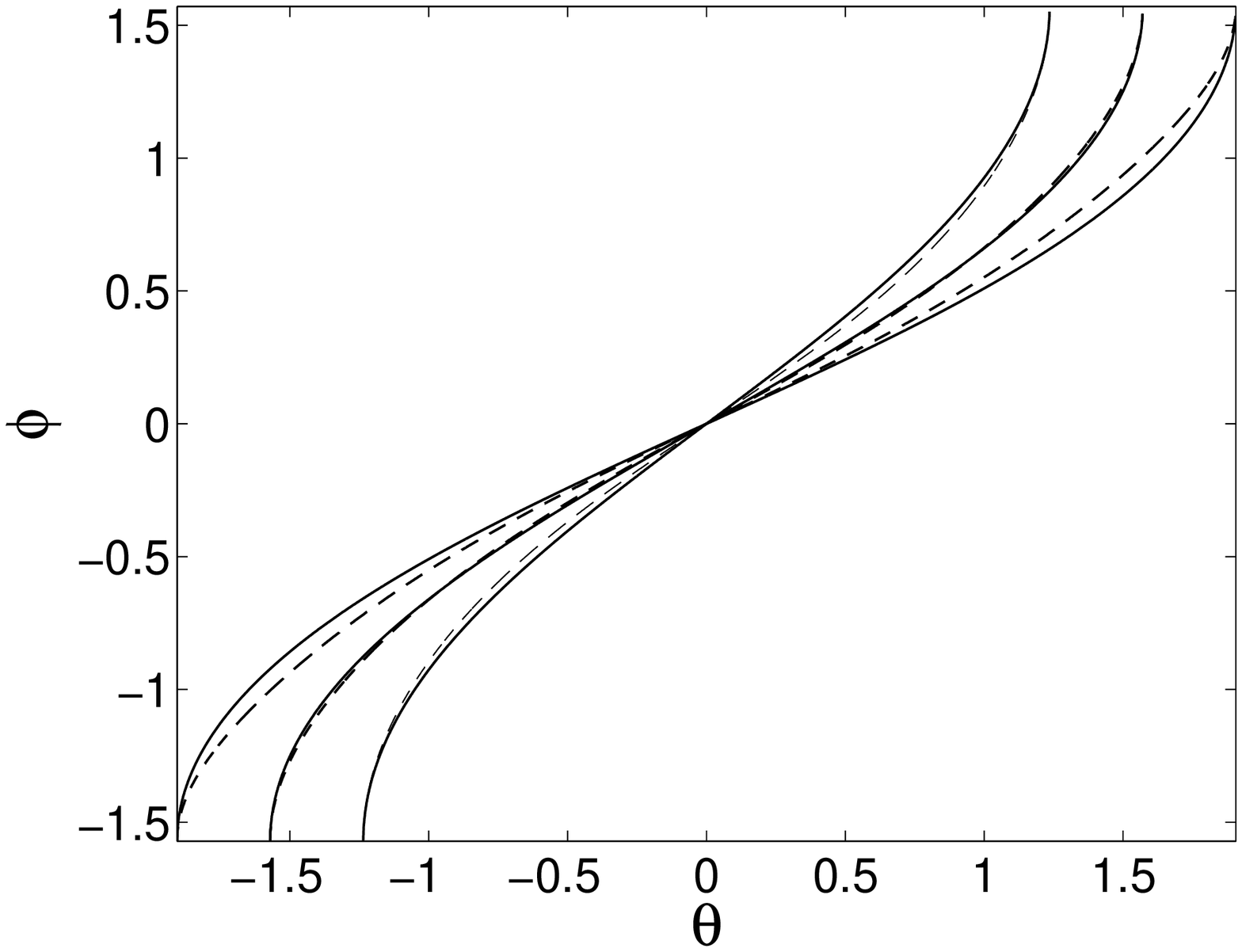}}}
 \caption{Left~: control term $f(\phi,t)$, at $t=0$. Right~: comparison of the exact and simplified changes of variables~: $\varphi$ is plotted as a function of $\theta$, for a given energy (and so for a given $J$). The plain line corresponds to the exact change, the dotted to the simplified one. The curves correspond, from bottom left to bottom right, to $E=0.5$, $0$ and $-0.5$.\label{fitang}}
\end{figure}

 Unfortunately, the exact change of variables from $(\varphi,J)$ to $(\theta,p)$ \cite{lichtenberg} has a singularity at the pendulum separatrices. In order to implement our control on the whole space, we use a simplified, but regular, change of variables which mimics the exact one in the region of the invariant torus predicted by the control (the error in this region is less than $4\%$) (see Fig.\ref{fitang}) :
\begin{equation*}
 \tan{\varphi} =  \frac{\theta}{p}.
\end{equation*}

 In terms of the $(\theta,p)$ variables, our control term now reads :
\begin{equation*} \tilde{f}(\theta,p,t)=f(\arctan{\frac{\theta}{p}},t), \end{equation*}
and its regularity is now the one of the function $(\theta,p) \mapsto \arctan \frac{\theta}{p} $.
Therefore, the controlled dynamics of a test-particle is given by Hamiltonian~:
\begin{equation*} H_c(\theta,p,t) = H_{1p}(\theta,p,t)+\tilde{f}(\theta,p,t). \end{equation*}
\begin{figure}[t] 
  \centerline{
    \mbox{\includegraphics[width=2.5in]{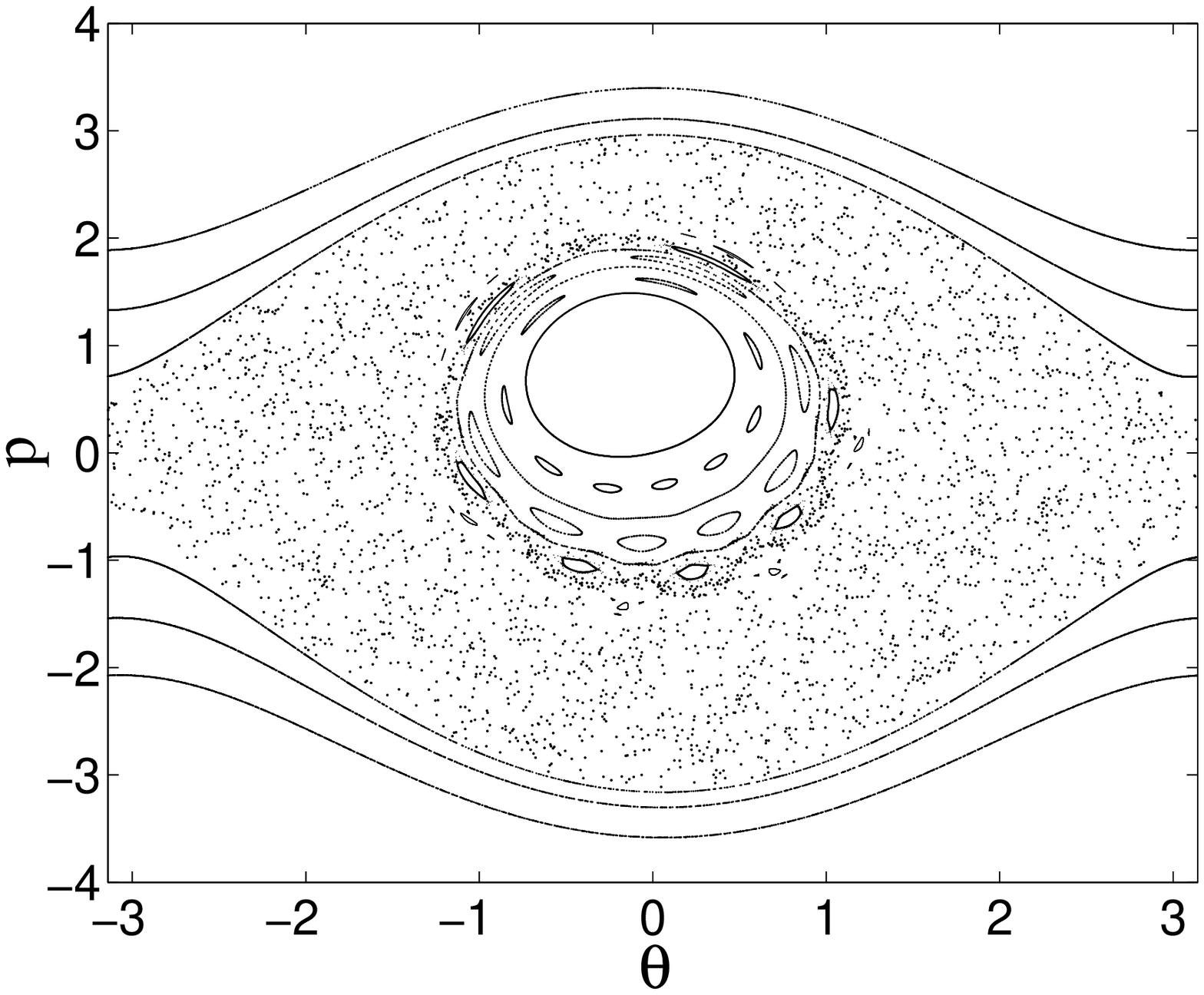}}
    \mbox{\includegraphics[width=2.5in]{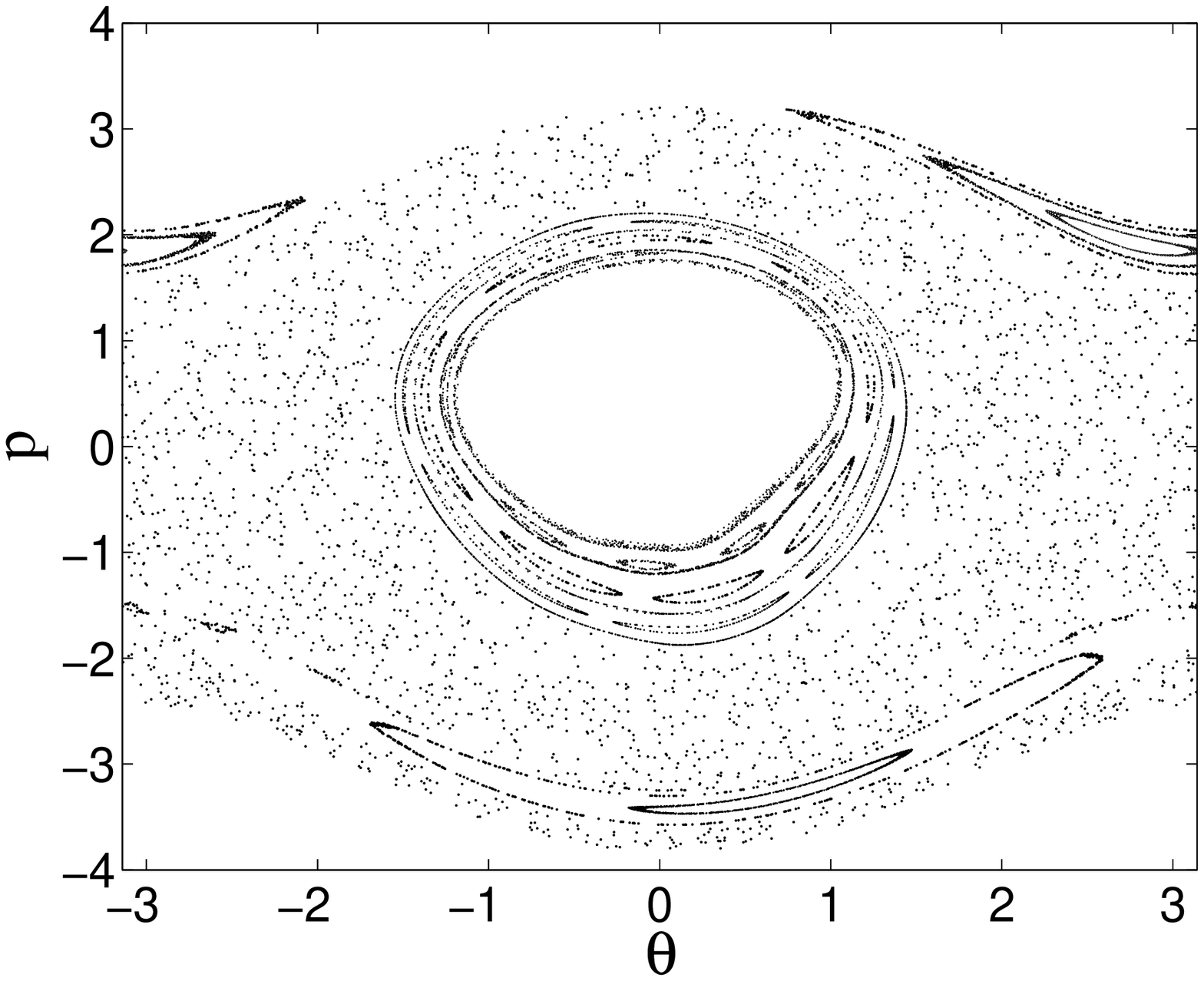}}}
  \caption{Poincaré sections of a test-particle of Hamiltonian $H_{1p}(\theta,p,t)$ (left) and $H_{1p}(\theta,p,t)+\tilde{f}(\theta,p,t)$ (right).\label{QP}}
\end{figure}
 In the $(\theta,p)$ variables (see Fig.\ref{QP}), the control is successful in reconstructing some invariant tori around the macro-particle. In other words, it enlarges this regular structure.
 
\section*{Conclusion}
In this paper, we considered a simplified mean-field approach to investigate the saturated dynamics of a Single Pass FEL. In particular, we showed 
that the size of the macro-particle can be increased by adding a small pertubation to the system, thus resulting in a 
low cost correction in term of energy. The main idea is to build invariant tori localized at specific positions~: this method is utterly general and 
could be succesfully used to adjust the size of the macro-particle, thus possibly enhancing the bunching factor. Analogously, by limiting the portion of phase-space spanned by the macro-particle, one could aim at   
stabilizing the laser signal. Two future lines of investigation are foreseen. First, we intend to implement the computed control term in the framework of the original
$N$-body self-consistent picture and explore possible beneficial effects for the evolution of the radiation field. Further, it is planned to apply the above technique 
to the case of the reduced Hamiltonian of \cite{tennyson,antoniazzi}. We point out that an experimental test of the control method on a modified Travelling Wave Tube has been done \cite{chandre} in absence of self-consistency.

\end{document}